\documentclass[conference]{IEEEtran}
\usepackage{soul}

\usepackage[cmex10]{amsmath}
\usepackage[utf8]{inputenc}

\usepackage{eurosym}
\usepackage{hyperref}

\usepackage{algorithm}
\usepackage{algpseudocode}
\usepackage{graphicx}
\usepackage{dsfont}

\usepackage{multirow}

\usepackage[utf8]{inputenc}
\usepackage{mathtools}
\usepackage[mathscr]{euscript}
\usepackage{amsmath}
\usepackage{amssymb}
\usepackage{amsfonts}
\usepackage{amssymb}
\usepackage{amsmath}
\usepackage{verbatim}
\usepackage[T1]{fontenc}
\usepackage[utf8]{inputenc}
\usepackage{latexsym}
\usepackage{enumerate}
\usepackage{indentfirst}
\usepackage{calligra}
\usepackage{ulem}
\usepackage{graphicx}
\usepackage{array}
\usepackage{float}
\usepackage{bbm}
\usepackage{color}

\usepackage{geometry}
\begin{document}
%
\title{Impact of subsidy on profitability of residential photovoltaics, battery and inverter in Flanders}

 \author{\IEEEauthorblockN{Md Umar Hashmi\IEEEauthorrefmark{1}, and
 Mohammed Meraj Alam\IEEEauthorrefmark{2}
 }
 \IEEEauthorblockA{\IEEEauthorrefmark{1} \textit{KU Leuven \& EnergyVille},
Genk, Belgium}
\IEEEauthorblockA{\IEEEauthorrefmark{2} FlexThor BV, Hasselt, Belgium}
 }

\maketitle

\begin{abstract}
The sustained future growth of renewable energy in the distribution network is governed by its financial viability.
The recent subsidies in Flanders for photovoltaic (PV) and battery are used to calculate the payback period on prosumer investment in residential PV, battery and inverter installations.
Realistic scenario of PV, battery and inverter are created based on the market catalogue of one manufacturer per product.
Using these scenarios, one-year simulations for a residential prosumer are conducted in a rolling horizon. Luminus, an energy service provider in Limburg,  time-of-use and feed-in-tariff contract rates are utilized.
The numerical evaluation indicates that the policies prefer new PV installations. The subsidy comparison for 2021 to 2024 is also conducted.
Oversizing PV compared to inverter size increases profitability.
Batteries performing only arbitrage may not attract much investment, as the payback period is high. 
This shows the need to maximize the value addition of prosumer investment on residential batteries by participating in  multiple goals.
\end{abstract}

\section{Introduction}
According to Belgian Solar National Association, the total installed capacity of Photovoltaic (PV) generation reached 6 GW by 2020 of which 900 MW have been added in 2020 despite the global pandemic \cite{link1}. 
Moreover, 74$\%$ of 6 GW installations took place in the Flemish part of Belgium (North), 23$\%$ in the Wallonia part of Belgium (South) and the rest in Brussels. With this trend, the renewable energy (PV + wind) production in the country increased by 31$\%$ and represents 18.6$\%$ of local electricity demand. In Flanders itself, this rapid growth in the PV installations took place in the residential sector which grew by 45$\%$ in 2020 as compared to 2019, primarily because of favourable regulatory framework \cite{link2}.

This framework states that the prosumers with installed PV base before 1st January 2021 are to be given a 15-year grace period before smart metering obligation comes into picture, which means that the prosumers can continue with the net metering or reverse counter \cite{link3}. However, these prosumers who want to continue with the legacy net metering are obliged to pay a yearly tax on the installed kVA rating of their inverter called \textit{prosumer tariff} roughly between 75 and 100 \euro  ~per year. On the other hand, for those prosumers who wish to install PV in or after 2021, they will receive a one-time subsidy on their installation with a mandatory upgrade to a smart meter (SM) at their premises \cite{link4}. 

Subsidies have already been introduced by the Flemish government on both the PV and battery system installations since 2021 and 2019, respectively, in order to boost renewable energy generation and local self-consumption. It has been observed that the prosumers in Flanders wish to invest in green technologies like PV + battery rather than stashing their money in the bank as the former gives far better return as compared to a bank. The Flemish ministry of power and energy has initiated the subsidy program to further achieve the target of 25$\%$ of renewable energy mix, raising the PV installations well beyond 11 GW by 2030. It is therefore strategically important in the eyes of government to scrap the reverse counter meters and plant smart meters to better monitor the development and achieve the said target \cite{link5}.

With reverse counter meter, broadly, a prosumer can choose a day meter with a single tariff or day and night meter with two different tariffs for day and night, respectively. Either a fixed price contract for 1 year or a dynamic price per quarter contract  can be selected by the prosumer. However, this comes with a yearly tax on the PV installation called \textit{prosumer tariff}. With smart meter, prosumers have two different tariffs, one for the consumption (high tariff) and another for the injection (low tariff) as the smart meter measures both the part separately. 

The recent trend in an unprecedented rise in PV + battery is adding a lot of pressure on the electricity network, which is forcing the network operators to promote self-consumption of PV power rather than injecting them into the network (using lower injection tariff and higher consumption tariff). This certainly requires a costly battery energy storage system that increases the upfront cost and payback period simultaneously. Since the full potential of PV + battery have not been harnessed yet, it becomes no longer attractive to the prosumers. There is a need among the prosumers to understand the return on investment, payback period and total upfront cost with and without subsidies.   

The key contributions of this paper are:\\
$\bullet $ Realistic PV, battery and inverter scenarios are created,\\
$\bullet $ Quantification of the impact of recent PV and battery subsidies in Flanders region in Belgium on their payback period on the prosumer investment for years 2021-24,\\
$\bullet $ Practical recommendations are provided for future investors planning to invest in residential scale PV, battery and inverter.

The paper is organized as follows. 
Section~\ref{section:scenarios} presents the scenario generation strategy for solar PV, battery and inverter.
Section~\ref{section3} summarizes the grid rules for prosumers in Flanders and subsidies new prosumers are eligible for.
Section~\ref{section4} compares the payback on investment on prosumer investment with different subsidy rules and
Section~\ref{section5} concludes the paper.

\section{Scenario generation: PV, battery and inverter}
\label{section:scenarios}
The problem of sizing solar PV and energy storage interfaced via a shared inverter is considered in this work.
Since batteries, inverters and PVs sizes are available in only limited discrete ratings instead of treating as continuous decision variables, they are used as input scenarios. For a given scenario which is a combination of PV, storage and inverter sizes, the prosumer simple payback period on their total investment and net savings are computed as the output. 
This output can be utilized by a consumer to select the best-suited combination of PV, storage and inverter.

In this work, we propose a mechanism for forming realistic combinations of (i) PV panel, (ii) battery, and (iii) inverter, which are evaluated for measuring the financial viability.
We consider the following assumptions in forming these scenarios:
\begin{itemize}
    \item We assume for only PV installation case, solar inverter is utilized. The solar inverter operates in one direction. In power electronic terms, diode bridge rectifier is utilized for solar inverter. On the other hand, hybrid inverters have bidirectional power flow capability so as it behaves both as an inverter and a rectifier.
    \item PV output and battery share a single bidirectional or hybrid inverter. Having a shared hybrid inverter is a common architecture used by many manufacturers. However, detailed evaluation of different architectures are beyond the scope of this work.
    \item We consider only 1 manufacturer's product line for PV panels, battery, solar inverter and hybrid inverter. The proposed framework can be extended for a larger array of products. However, for computational tractability of the proposed framework, only one manufacturer is considered.
    \item In order to generate realistic scenarios, the inverter size should be between 0.55 to 1.1 times the sum of peak power output of PV and/or battery. This range captures oversizing and under sizing of inverters.
    \item For instances where PV size plus battery power output is lower than 0.55 times the lowest inverter size, in such cases the smallest inverter is selected to form a scenario. Similarly, for instances where PV size plus battery power output is greater than 1.1 times the highest inverter size, in such cases the largest inverter is selected to form a scenario.
    \item The batteries are assumed to be 1C-1C\footnote{Battery xC-yC denotes battery takes 1/x hours to fully charge and 1/y hours to fully discharge. 1C-1C denotes battery takes at least 1 hour to charge (or discharge) from fully discharged (or full charged) state to completely charged (or completely discharged) state.} type. The charging and discharging efficiency (given as $\eta_{\text{ch}}, \eta_{\text{dis}}$) is assumed to be 95\%.
    The minimum permissible battery charge level is denoted as $b_{\min}$ is 10\% of maximum permissible level $b_{\max}$.
    \item The inverter (solar and hybrid) is assumed to have an operating efficiency of 98\%.
\end{itemize}
Next, we detail the product ratings and associated cost used for scenario generation.

\textbf{LG solar panels}:
LG is one of the leading manufacturer of solar panels.
Table \ref{tab:solarLG} lists the monocrystalline LG solar panels.
The first column tabulates different sizes of panels in kWp which can be combined in series or parallel constrained by roof area, panel voltage considerations; if the panel voltage does not match that of inverter input specifications then a different DC-AC inverter would be required, this would affect the system cost.
In this work, we consider panels of size 0.325 kWp as its per kWp cost is the least. Therefore, the possible solar sizes will be integral multiples of 0.325 kWp.
The peak installed solar is assumed to be 6 kWp. Thus, there are 19 PV combinations.
\begin{table}[ht]
 \footnotesize
	\caption {Price comparison of Monocrystalline LG solar panels} \vspace{-10pt}
	\label{tab:solarLG}
	\begin{center}
		\begin{tabular}{c|c|c}
			\hline
			{\textbf{Panel peak power in kWp}}& {\textbf{price in \euro /panel}} & {\textbf{\euro /kWh}} \\
			\hline 
			\hline
			0.300 &  172 & 860\\
             \hline 
            \hl{0.325} &  174 & 535\\ 
             \hline
            0.350 &  211 & 844\\ 
             \hline
            0.370 &  297 & 802\\ 
             \hline
            0.400 &  249 & 622\\
            \hline
        \end{tabular}
		\hfill\
	\end{center}
\end{table}

\textbf{LG Chem battery}: This manufacturer has four battery sizes which can be used for LV residential consumer application, compared to Tesla with only 2 products. 
Table \ref{tab:lgchembatte} lists LG Chem 48 V batteries which could be used for residential applications.

\begin{table}[ht]
 \footnotesize
	\caption {Price comparison of 48 Volt LG Chem home battery storage} \vspace{-10pt}
	\label{tab:lgchembatte}
	\begin{center}
		\begin{tabular}{c|c|c}
			\hline
			{\textbf{Battery capacity in kWh}} &   {\textbf{price in \euro}} & {\textbf{Maximum charging current}} \\
			\hline 
			\hline
			3.3 &  2349 & 71.4\\
             \hline 
            6.5 &  3409 & 100\\ 
             \hline
            9.8 & 4438 & 119\\ 
             \hline
            13.1 &  5590 & 119\\ 
            \hline
        \end{tabular}
		\hfill\
	\end{center}
\end{table}

\textbf{SMA solar inverters}: It is typical for LV prosumers needs to abide by inverter size rules. Typically, single phase inverters cannot exceed 6 kVA.
Table \ref{tab:solarinverter} lists SMA solar inverters.

\begin{table}[ht]
 \footnotesize
	\caption {Price comparison of SMA grid-connected 1-phase inverter} \vspace{-5pt}
	\label{tab:solarinverter}
	\begin{center}
		\begin{tabular}{c|c|c}
			\hline
			Index & {\textbf{kVA rating}} &  {\textbf{price in \euro}}   \\
			\hline
			\hline 
			1&1.5 & 539 \\
             \hline 
            2&2 & 647  \\ 
             \hline
            3&2.5 & 729 \\ 
             \hline
            4&3 & 875 \\ 
             \hline
            5&3.6 & 929  \\
            \hline
            6&4 & 979  \\
            \hline
            7&5 & 1049  \\
            \hline
            \multicolumn{3}{|c|}{www.europe-solarstore.com accessed on 14 March 2021} \\
            \hline
		\end{tabular}
		\hfill\
	\end{center}
\end{table}

\textbf{Solaredge hybrid inverters}: Solaredge is a leading manufacturer of inverters.
We tabulate Solaredge hybrid inverters in Table \ref{tab:hybridinverter}.

\begin{table}[ht]
 \footnotesize
	\caption {Price comparison of Solaredge hybrid 1-phase inverters} \vspace{-10pt}
	\label{tab:hybridinverter}
	\begin{center}
		\begin{tabular}{c|c|c}
			\hline
			Index & {\textbf{kVA rating}} & \textbf{price in \euro}  \\
			\hline
			\hline 
			1 & 2.2 & 1529 \\
             \hline 
            2 & 3 & 1599 \\ 
             \hline
            3 & 3.5 & 1689 \\ 
             \hline
            4 & 3.68 & 1699 \\ 
             \hline
            5 & 4 & 1739 \\
            \hline
            6 & 5 & 1779 \\
            \hline
            7 & 6 & 1929 \\
            \hline
            \multicolumn{3}{|c|}{www.europe-solarstore.com accessed on 17 March 2021} \\
            \hline
		\end{tabular}
		\hfill\
	\end{center}
\end{table}

Taking the above assumptions into consideration, we reduce the total number of scenarios from worst-case number of scenarios from 665 to 185, a reduction of $\approx 71.9\%$. 665 combinations are formed by 19 PV types (including 0 PV size), 5 battery types (including no battery) and 7 inverter types.

\begin{figure}[!htbp]
	\center
	\includegraphics[width=3.6in]{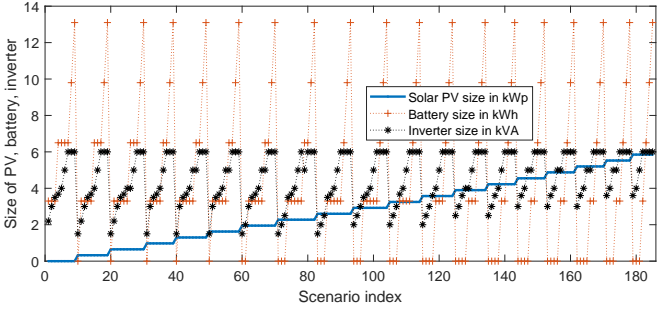}
	\vspace{-12pt}
	\caption{\small{Generated scenarios for solar pv, battery and inverters}}
	\label{fig:scenarios}
\end{figure}

\begin{figure}[!htbp]
	\center
	\includegraphics[width=3.5in]{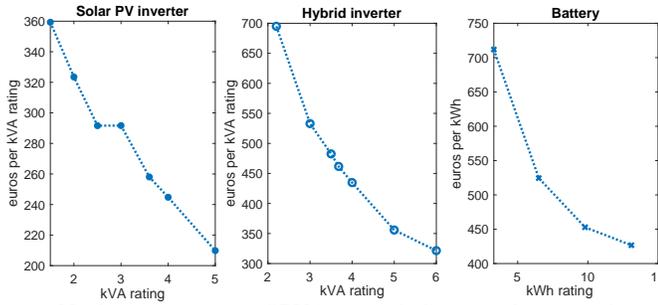}
	\vspace{-10pt}
	\caption{\small{Nonlinearity in cost of PV inverter, bidirectional inverter, battery. The cost tends to decrease with increased size of inverter and battery.}}
	\label{fig:market}
\end{figure}

Fig.~\ref{fig:scenarios} shows 185 scenarios. Note the solar and hybrid inverters are not distinguished in Fig.~\ref{fig:scenarios}. However, from Table \ref{tab:solarinverter}, Table \ref{tab:hybridinverter} and Fig.\ref{fig:market} we observe:
(a) the nonlinear growth of per unit cost of battery, solar and hybrid inverters,
(b) solar inverter costs less than 50\% of the cost of hybrid inverters,
(c) there is an incentive in oversizing as per capita cost dips with size.

\section{Grid rules for prosumers in Belgium}
\label{section3}
Belgian residential electricity contracts provided by Luminus are used for techno-economic analysis for PV, battery and inverter sizing.
Luminus offers many variants of consumer contracts such as \textit{ComfyFlex Pro}, \textit{Comfy Pro+}, \textit{Actief+ Elektriciteit}, to name a few \cite{luminus_elektriciteit}.
These consumer contracts are settled every three months (at the end of the period). A forecast of tariffs for the next three months are provided. For some of these contracts the tariffs can vary by more than 50\%. For example, the day tariff for \textit{Basic Elektriciteit} during January to March 2021 was 6.32 \euro cents/kWh and the forecast day tariff for October to December 2021 is expected to be 11.78 \euro cents/kWh.
In this work, we utilize consumer contract \textit{Actief+ Elektriciteit }for the month of January to March 2021 \cite{luminus_actief}.
The four billing mechanism under this contract are listed in Table \ref{tab:consumerContracts}.
The billing of electricity has three main components:
(a) cost of consumption and injection of electricity,
(b) transportation of electricity, and
(c) government and value added taxes.
For this work, we are interested in variable component of electricity consumption cost and prosumer tax (\textit{Prosumententarief}) as detailed in Table \ref{tab:consumerContracts}.
The prosumer tax is proportional to the size of inverter capacity. Larger the size of inverter, greater will be prosumer tax.
Prosumer tax is region dependent. For example, in Limburg the prosumer tax is 72.25 \euro per year per kW which is lowest in Flanders, on the other hand Gaselwest under Fluvius charges 112.42 \euro per year per kW.
For this work we use contracts C3 as detailed in Table \ref{tab:consumerContracts}. C3 corresponds to consumers with smart meter with 2 level time-of-use (ToU) with different buying and selling price of electricity. Prosumer tax is not levied upon C3.
C4 corresponds to consumers without SM. It also has 2 level ToU tariff with equal buying and selling prices with prosumer tax.
{In Table \ref{tab:consumerContracts}, the night charge is applied for 10 pm to 7 am}

\begin{table*}[!htbp]
 \footnotesize
	\caption {Some residential consumer contracts offered by Luminus in Flemish Region} \vspace{-10pt}
	\label{tab:consumerContracts}
	\begin{center}
		\begin{tabular}{p{0.25cm}|p{1.99cm}||p{1cm}|p{1.3cm}|p{1.65cm}|p{2.2cm}|p{1.2cm}|p{1.8cm}| p{2.29cm}}
			\hline
			\multirow{2}{*}{\textbf{Id}} &  \multirow{2}{*}{\textbf{Contract}} & \textbf{Fixed charge \euro/year}& \textbf{Day charge c\euro/kWh} & \textbf{Night charge}  \textbf{c\euro / kWh}& \textbf{Injection rate c\euro/kWh}& \textbf{Prosumer tax}& \textbf{Fixed distribution cost} & \textbf{Variable distribution cost}  \\
			\hline 
			\hline
			C1. & Single rate meter (\textit{\underline{with}} SM) & 90.75 \euro/yr & \multicolumn{2}{|c|}{7.98 c\euro/kWh} & 3.21 c\euro/kWh & No & 13.64 \euro/yr+ 0.43\euro/month & 9.04+2.10+ 0.3508 c\euro/kWh\\
             \hline 
             \hline
             C2. & Single rate meter (\textit{\underline{without}} SM) & 90.75 \euro/yr & \multicolumn{2}{|c|}{7.98 c\euro/kWh} & - & 72.25 \euro/kW/yr& 13.64 \euro/yr+ 0.43\euro/month & 9.04+2.10+ 0.3508 c\euro/kWh\\ 
             \hline
             \hline
             \hl{C3.} & Dual rate meter (\textit{\underline{with}} SM) & 90.75 \euro/yr & {8.6 c\euro/kWh} & {5.72~c\euro/kWh} (also weekend \& holidays) & 3.59c\euro/kWh (day) \& 2.11c\euro/kWh (night) & No & 13.64 \euro/yr+ 0.43\euro/month & 9.04~(day) + 7.23~(night) +2.10 +0.3508 c\euro/kWh\\ 
             \hline
             \hline
             {C4.} & Dual rate meter (\textit{\underline{without}} SM) & 90.75 \euro/yr & {8.6 c\euro/kWh} & {5.72~c\euro/kWh} (also weekend \& holidays) & 8.6 c\euro/kWh (day), \& 5.72c\euro/kWh (night) & 72.25 \euro/kW/yr & 13.64\euro/yr+ 0.43\euro/month & 9.04~(day) + 7.23~(night) +2.10 +0.3508 c\euro/kWh\\ 
             \hline
		\end{tabular}
		\hfill\
	\end{center}
\end{table*}


\subsection{Subsidy for PV installation in Flanders}
With the introduction of smart meter in the Flemish region, there is no longer a reversing meter for those who start using solar panels on their rooftop from 1 January 2021.In addition to this, anyone who installs new solar panels from 1 January 2021 is eligible for a one-time subsidy from the distribution system operator Fluvius. However, one has to meet all the conditions to be eligible for this subsidy \cite{vlaanderen_solar}.

\begin{itemize}
    \item The subsidy amount is adjusted on 1st of January every year, which is determined by the date of commissioning of the solar panels.
    \item The subsidy amount is limited to a maximum of 40\% of the investment costs, including VAT.
    \item This amount of subsidy is independent of a home connected to the distribution network.
    \item The subsidy is calculated based on the power of the solar panels themselves and not the power rating of the inverter.
\end{itemize}
\begin{table}[!htbp]
 \footnotesize
	\caption {Subsidy on PV panels } \vspace{-10pt}
	\label{tab:subsidyPV}
	\begin{center}
		\begin{tabular}{c|c|c|c}
			\hline
			Year & {\textbf{$\leq$ 4 kWp}} & \textbf{$\geq$ 4 kWp and $\leq$ 6 kWp} & \textbf{Maximum subsidy } \\[1mm]
			\hline 
			2021 & 300 \euro/kWp & 150 \euro/kWp & 1500 \euro\\
             \hline 
            2022 & 225 \euro/kWp & 112.5 \euro/kWp & 1125 \euro\\ 
             \hline
            2023 & 150 \euro/kWp & 75 \euro/kWp & 750 \euro\\ 
             \hline
            2024 & 75 \euro/kWp & 37.5 \euro/kWp & 375 \euro\\ 
             \hline
            
		\end{tabular}
		\hfill\
	\end{center}
\end{table}
The subsidy rules on PV are summarized in Table \ref{tab:subsidyPV}. For example, in 2021, the subsidy is 300 \euro/kWp for solar installations up to a maximum of 4 kWp and an additional 150 \euro/kWp from 4 kWp to 6 kWp of installations, which corresponds to a maximum of 1500 \euro.

\subsection{Subsidy on home batteries in Flanders}
As presented in Table \ref{tab:subsidyBattery}, the subsidy on batteries applied by the Flemish government in 2021 is 300 Euros per kWh capacity of the battery. This is applicable on the Lithium and gel-based battery type and not on the lead-acid battery. The bigger is the capacity of the battery, higher is the subsidy up to 9 kWh. If the battery size is more than 9 kWh, then no additional subsidy is applicable on the capacity exceeding 9 kWh.

\begin{table}[!htbp]
 \footnotesize
	\caption {Subsidy on Home batteries in \euro per kWh \cite{vlaanderen_battery}} \vspace{-10pt}
	\label{tab:subsidyBattery}
	\begin{center}
		\begin{tabular}{c|c|c|c|c}
			\hline
			Battery capacity & \textbf{2021} & \textbf{2022} & \textbf{2023} & \textbf{2024}\\[1mm]
			\hline 
			\hline
			0 to 4 kWh & 300  & 225 & 150 & 75 \\
             \hline 
            4 to 6 kWh & 300  & 187.5  & 125  & 62.5 \\ 
             \hline
            6 to 9 kWh &  250  & 150  & 100  & 50 \\ 
             \hline
            9 kWh onward & \multicolumn{4}{|c|}{No additional amount}   \\ 
             \hline
            Maximum subsidy & 2550 \euro  & 1725 \euro & 1150 \euro & 575 \euro \\ 
             \hline
		\end{tabular}
		\hfill\
	\end{center}
\end{table}

\subsection{Initial investment cost after subsidies}
\begin{figure*}[!htbp]
	\center
	\includegraphics[width=6.8in]{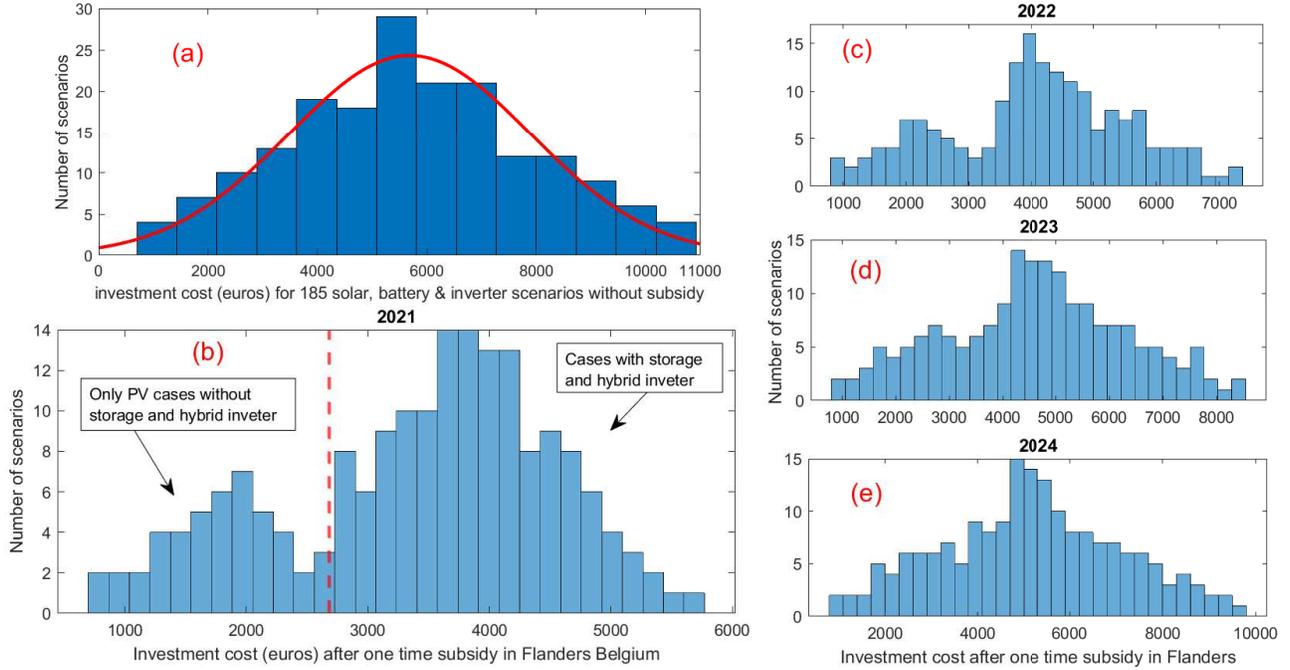}
	\vspace{-3 pt}
	\caption{\small{The distribution of initial investment for the scenarios without and with subsidy in Flanders, Belgium.}}
	\label{fig:distribution}
\end{figure*}

In Section \ref{section:scenarios} we identify 185 scenarios of solar PV, battery and inverter.
We apply the government subsidies on the market cost.
Figure \ref{fig:distribution} shows that how subsidies incentivize solar PV installations over batteries.
Note that this disparity is also caused due to high cost of hybrid inverters compared to solar inverters.

In order to increase profitability of batteries, co-optimization of multiple goals maybe needed \cite{hashmi2019co}.

\section{Payback on investment for a prosumer}
\label{section4}
The simple payback period on prosumer investment on PV, battery and inverter are calculated for the scenarios described in Section \ref{section:scenarios}. The Belgian consumer profiles for one prosumer is used from Linear database described in \cite{dupont2012linear}.
For this analysis, the payback period is calculated for contract C3, described in Table \ref{tab:consumerContracts}.
The local optimization of battery is based on the algorithm proposed in \cite{hashmi2019optimal, hashmi2019optimization}.
The optimization is performed in a rolling horizon of one day.
The net cost of consumption is aggregated over a year. The annual savings is calculated as the difference of the cost of consumption in absence and with PV, battery and inverter.

\begin{figure}[!htbp]
	\center
	\includegraphics[width=3.2in]{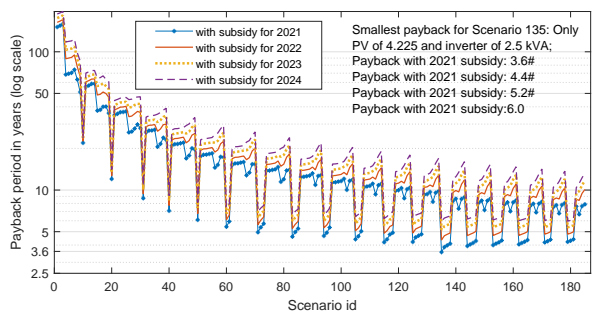}
	\vspace{-8 pt}
	\caption{\small{Payback period on log-scale for 185 scenarios of PV, battery and inverter with Flanders subsidy for year 2021-2024 for contract C3.}}
	\label{fig:payback}
\end{figure}

\begin{figure}[!htbp]
	\center
	\includegraphics[width=3.1in]{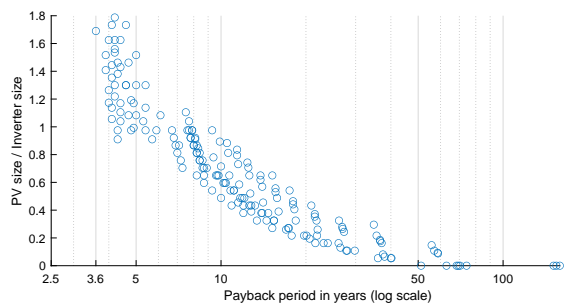}
	\vspace{-8 pt}
	\caption{\small{Relationship between the ratio of PV and inverter size and payback period.}}
	\label{fig:pvratio}
\end{figure}

\begin{figure}[!htbp]
	\center
	\includegraphics[width=3.2in]{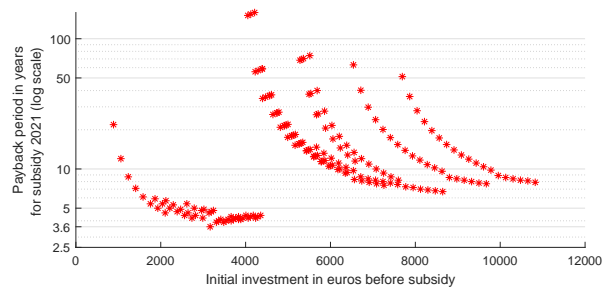}
	\vspace{-8 pt}
	\caption{\small{Initial investment and payback period.}}
	\label{fig:investmentpay}
\end{figure}

Fig.~\ref{fig:payback} shows the simple payback period for the prosumer based on extrapolation of one year of simulation results for 185 scenarios for subsidy rates for 2021.
Observe that some scenarios have a very high payback period. The initial scenarios denote a small battery with a hybrid inverter. The payback on such an investment exceeds 100 years.
Fig.~\ref{fig:pvratio} shows the negative correlationship between PV and inverter size ratio and the payback period.
Therefore, oversizing PV with respect to inverter makes lots of sense for a prosumer.
Fig.~\ref{fig:payback} highlights that for scenario 135, which consists of only PV of size 4.225 and inverter size of 2.5 kVA leads to payback period of 3.6, 4.4, 5.2 and 6.0 years with subsidy rates for 2021, 2022, 2023, and 2024 respectively.
Fig.~\ref{fig:investmentpay} shows the relation of initial investment (without subsidy) and payback period.
Only solar cases without battery have typically a low payback period. This implies that the residential batteries only performing arbitrage may not lead to financial feasibility.

The key findings from this numerical study are as follows:
\begin{itemize}
    \item 49.26\% increase in payback period for PV, battery and inverter scenarios for subsidies for 2024 compared to 2021,
\item oversizing PV and under sizing inverter without storage leads to reduction in payback period of up to 3.6 years for 2021 Flanders subsidy.
For such cases, the ratio of PV size over inverter size is approximately 1.5,
\item storage subsidies are not as significant as PV subsidies in Flanders further, inclusion of a hybrid inverter leads to reduced profitability of batteries,
\item profitability of battery increases for a larger battery compared to smaller battery. This is because of flattening of cost of large inverters and battery.
\end{itemize}

\section{Conclusion}
\label{section5}
The recently introduced subsidies on PV and battery in Flanders in Belgium are evaluated in this paper. 
Based on realistic scenarios of PV, battery and inverter based on one manufacturer product for each product, one year of cost of electricity consumption is calculated.
The profit is defined as the net savings due to the installation of PV, battery and inverter. This prosumer saving is used to calculate simple payback period on prosumer investment, considering subsidies.
We observe that early mover has several advantages, which decreases on average around 50\% by the year 2024.
We also observe, Flanders subsidy policy favour PV installations over battery, which is evident from only PV installations having a payback period of as low as 3.6 years. However, most profitable scenario with battery has a simple payback period of 7 or more years.
This implies installations of batteries should consider additional revenue streams for ensuring financial feasibility.

\bibliographystyle{IEEEtran}
\bibliography{reference}

\end{document}